\documentclass[12pt]{article}
\usepackage{epsfig}
\begin{document}
\begin{center}
{\Large {\bf Quantization of empty space. \\

}} \vskip-40mm \rightline{\small ITEP-LAT 2005-23} \vskip 30mm

{
\vspace{1cm}
{ M.A.~Zubkov$^a$  }\\

 \vspace{.5cm} { \it
$^a$ ITEP, B.Cheremushkinskaya 25, Moscow, 117259, Russia }}
\end{center}
\begin{abstract}
We suggest to use "minimal" choice of quantum gravity theory, that is the
quantum field theory, in which  space-time is seen as Riemannian space and
metric (or vierbein field) is the dynamical variable. We then suggest to use
the simplest acceptable action, that is the squared curvature action. The
correspondent model is renormalizable, has the correct classical limit without
matter and can be explored using Euclidian path integral formalism. In order to
get nonperturbative results one has to put this model on the lattice. While
doing so serious problems with  measure over dynamical variables are
encountered, which were not solved until present. We suggest to solve them
using the representation of Riemannian space as a limiting case of Riemann -
Cartan space, where the Poincare group connection plays the role of dynamical
variable. We construct manifestly gauge invariant discretization of Riemann -
Cartan space. Lattice realization of Poincare gauge transformation naturally
acts on the dynamical variables of the constructed discretization. There exists
local measure invariant under this gauge transformation, which could be used as
a basic element of lattice path integral methods. The correspondent lattice
model appears to be useful for numerical simulations.
\end{abstract}

\today \hspace{1ex}

\newpage

\section{Introduction}

Quantum theory of gravity should describe behavior of the Early Universe.
Information about its structure comes mostly from astrophysical observations.
Unfortunately present data does not allow to make a definite choice between
different approaches to construction of quantum gravity theory.  Therefore we
suggest to make a choice, which we would call "minimal". Namely, first we fix a
functional integral method of quantization and do not consider Hamiltonian
methods.   Next, instead of using complicated mathematical structures we
consider Riemannian manyfold, which is used in general relativity. Therefore
our dynamical variable is the field of metric. Then, there are two ingredients
of the theory to be chosen: the action and the measure. In \cite{TELEPARALLEL}
we motivated that the squared curvature action is the most appropriate choice.
In the next section we briefly repeat our arguments. It seems that this model
should describe pure gravity without matter very naturally. And even if the
correct theory is different, the constructed one may serve as a low energy
approximation. Unfortunately, inclusion of matter is not straightforward. And
in spite of we guess, that it could be done, the present paper is called
"quantization of empty space".

When trying to use the chosen theory  in practice, we have to put it on the
lattice, which is the only way to explore it nonperturbatively. It is also the
only way to define  measure over dynamical variables. Here we encounter serious
problems, which are related to the wideness of the gauge group (which is the
group of general coordinate transformations). Namely, in the naive
discretizations using rectangular lattices gauge invariance is lost. It was
shown, that in ordinary gauge theories the discretization, which is not gauge
invariant, is not appropriate (say, in nonabelian models there is no
confinement of fundamental charges \cite{NONABELIAN}). On the other hand, in
Regge calculus, which is gauge invariant discretization of Riemannian manyfold,
the only measure whose appearance is motivated by reasonable (although not
perfect) arguments is Lund - Regge measure, which corresponds to a certain
metric\footnote{This metric is induced by the simplest possible continuum
metric on the space of Riemannian geometries.}. This measure is nonlocal and so
complicated, that practical numerical simulations seem to be impossible
\cite{AMBJORN}. In \cite{REGGE2D} the way to solve this problem in two
dimensions was suggested. In this approach partially gauge fixed version of the
theory is used. So, we obtain local and simple measure and instead loose a
small part of gauge invariance (its conformal part). However, even this
semi-solution can not be expanded to physically interesting case of four
dimensions.

Alternative approach to discretization of Riemannian space was suggested in
\cite{TELEPARALLEL}, where Riemannian manyfold of simple topology is seen as
Weitzenbock space. (Riemannian manyfold and Weitzenbock space are just
different ways to look at the same entirety.) In section $3$ we remind this
construction. In this discretization the inverse vierbein, i.e. a translational
connection plays the role of fundamental variable. Measure over this variable
 is local. Negative feature of this discretization is that
the action depends on Riemannian curvature, which should be   expressed through
the inverse vierbein in a nontrivial way. In \cite{TELEPARALLEL} there were
suggested to use the approximate expression, which tends to the correct one in
the naive continuum limit. Thus, gauge invariance there is not exact. But even
with this expression the model is not useful for numerical simulation as any
discretization of the model with second derivatives in the action.

Therefore in this paper we suggest to apply another approach. We consider
 Riemann-Cartan space, in which both $O(4)$
 connection\footnote{Here and further in the text we imply, that Wick rotation to Euclidean
 signature is performed. So, we deal with Euclidean path integral formalism.}
  and the inverse vierbein are
the dynamical variables. We do not require vanishing of torsion (which would
lead to appearance of Riemannian manyfold) or vanishing of curvature (which
would lead to appearance of Weitzenbock space). Instead we construct manifestly
gauge invariant discretization of Riemann-Cartan space and suppose, that
vanishing of torsion should be achieved dynamically. Elements of Poincare group
that is the gauge group of the theory naturally act on the dynamical variables
of the discretized model. The construction is described in section $4$. In
section $5$ we consider the action of the model. It is worth mentioning, that
discretization of Poincare gravity, which is based on the Riemann-Cartan space,
has already been considered in literature (see, for example, \cite{unitarity}).
However, the correspondent constructions were not manifestly gauge invariant as
Regge discretization of Riemannian space or as the discretization suggested in
this paper. Moreover, in \cite{unitarity} and related publications Poincare
group was not the real gauge group of the lattice models.

In section $6$ we discuss problems with measure encountered in Regge
discretization, where  measure is constructed in order to correspond to the
metric on space of Riemannian geometries. We demonstrate, using the finite
dimensional analogy, that Lund - Regge measure may not be appropriate at all.
Therefore, we construct measure over dynamical variables in the suggested
lattice model using another approach. We use symmetry properties of the
continuum model (that is a lattice gauge theory, in which Poincare group plays
the role of the gauge group). Gauge invariance of the measure is explored in
order to determine it. This construction is considered in section $7$. In
section $8$ we end with our conclusions.

\section{"Minimal" quantum gravity}

The simplest choice of the action would be Einstein-Hilbert action. However,
the resulting theory appears to be nonrenormalizable. Moreover, after Wick
rotation to Euclidian signature the action becomes unbounded below. This
indicates that the vacuum state of the model should be highly curved with
fractal dimension far from its physical value $4$.

It is well - known, however, that ultraviolet divergences in  quantum gravity
with the action that contains  additional squared curvature term can be
absorbed by an appropriate renormalization of the coupling constants
\cite{renormalizable}.  The model considered  in \cite{renormalizable} and
related publications has the following action:
\begin{equation}
S = \int \{ \alpha (R_{AB} R_{AB} - \frac{1}{3}R^2) + \beta R^2 - \gamma m_p^2
R + \lambda m_P^4\} |E| d^4 x, \label{S1}
\end{equation}
where $|E| = {\rm det} E^A_{\mu}$, $E^A_{\mu}$ is the inverse vierbein, the
tetrad components of Ricci tensor are denoted by $R_{AB}$, and $R$ is the
scalar curvature. Coupling constants $\alpha, \beta, \gamma$ and $\lambda$ are
dimensionless while $m_p$ is a dimensional parameter. Linearized theory (around
flat background) contains graviton together with additional tensor and scalar
excitations. The whole propagator behaves like $\frac{1}{q^4}$ in ultraviolet
while $\alpha, \beta \ne 0$. Tensor excitation is a ghost, which leads to  loss
of unitarity.

The requirement that the action (\ref{S1}) is bounded below leads to appearance
of a tachyon. This indicates that flat space is not  real vacuum of the model.
The tachyon would disappear if we construct  perturbation expansion around the
background that minimizes (\ref{S1}) \cite{a_free}. In addition to the
ultraviolet divergences the perturbation expansion may also contain infrared
divergences. In order to separate their consideration from the consideration of
the ultraviolet ones we have to use an additional regularization. This can be
done, for example, if the invariant volume $V$ of the space - time manyfold is
kept constant. Then after usual regularization (say, the dimensional or lattice
regularization) is removed and all the ultraviolet divergences are absorbed by
redefinition of the coupling constants, each term of the perturbation expansion
is finite. In the theory with fixed invariant volume  cosmological constant
does not influence the dynamics and the action is bounded  below if ($\alpha
\ge 0$, $\beta > 0$, $\gamma\ne 0$) or ($\alpha \ge 0$, $\beta \ge 0$, $\gamma
= 0$). The renormalization group analysis shows \cite{a_free} that at $\alpha,
\beta >0$ there exists a region of couplings such that the theory is asymptotic
free in $\alpha$ and $\beta$ while $\gamma$ can be made constant (up to one -
loop approximation). Divergences may appear also in the limit $V \rightarrow
\infty$. However, similar divergences  appear in QED but they are compensated
by ejection of soft photons. Probably, the same mechanism may work here as
well.

Unfortunately classical Newtonian limit cannot be obtained directly from the
action (\ref{S1}) unless it is not bounded from below. However, if we start
from pure gravity model with the action (\ref{S1}) (with $\lambda = 0$) and
rotate it back to Minkowski signature, solutions of Einstein equations would
satisfy the appeared classical equations of motion. They are not the only
solutions of the equations of motion. However, at $\gamma = \lambda = 0$
Einstein spaces minimize the (Euclidean) action\footnote{And tachyon
disappears.}. Therefore it could be interesting to consider the theory with the
action (\ref{S1}) such that at some scale the {\it renormalized} couplings
$\gamma$ and $\lambda$ vanish.  Then massive point - like objects could be
treated,  as space - time singularities \cite{mass_singularities} and the
Newtonian limit appears as an asymptotic of black hole solutions. Suppose that
the line - like singularity is embedded into the space - time. Then Einstein
equations in empty space would lead to Einstein equations in the presence of a
particle moving along the mentioned singularity. Its mass is not fixed by the
field equations but it is proved to be constant along the world trajectory
\cite{mass_singularities}. This indicates that
 matter can be introduced into quantum gravity
theory with the action (\ref{S1}) in such a way that it reproduces general
relativity at $\alpha > 0,\, \beta > 0, \lambda = 0$.

\section{Discretization of  Weitzenbock space.   }

This approach is based upon teleparallel formulation of general relativity
\cite{teleparallel}. The correspondent geometrical construction is the so -
called Weitzenbock space that appears as a limiting case of a more general
concept - Riemann - Cartan space. The latter is a tangent bundle equipped with
the connection from Poincare algebra. Poincare group consists of $O(4)$
rotations and translations\footnote{Here and below we always remember, that
Wick rotation to Euclidean signature is performed. Nevertheless for simplicity
we shall call the group, which consists of $O(4)$ rotations and translations as
Poincare group keeping in mind that back rotation to Lorentzian signature
should be done after calculations using Euclidean path integral methods are
performed.}. Translational part of connection can be identified with the
inverse vierbein and defines space - time metric. The correspondent part of the
curvature becomes  torsion. Riemannian geometry appears when torsion is set to
zero. Weitzenbock geometry is an opposite limit: $O(4)$ part of Poincare
curvature is set to zero while torsion remains arbitrary. Teleparallel gravity
is the theory of Weitzenbock geometry, i.e. a translational gauge theory.

If the space - time manyfold is parallelizable  zero curvature $O(4)$
connection can be chosen equal to zero. Therefore the only dynamical variable
is the inverse vierbein, treated as a translational connection. Usually action
in teleparallel gravity is expressed through the translational curvature
(torsion); space - time and internal indices can be contracted by the vierbein.
The equivalence between continuum theories of Riemannian and Weitzenbock
geometries can be set up if everything in Riemannian geometry is expressed
through the inverse vierbein and the latter is identified with the
translational connection.

Our discretized space is composed of flat pieces connected together. We
consider two cases: when each such piece has the form of a simplex or when it
has the hypercubic form. So, we talk about either simplicial or hypercubic
lattice. For the further convenience we refer to simplices (hypercubes) as to
elements of the lattice.
  Form of the
lattice elements is fixed by the set of vectors ${\bf e}_{\mu}$ connecting the
center of the element with its vertices. The expression of ${\bf e}_{\mu}$
through elements of the orthonormal frame ${\bf f}_A$ ($A = 1,2,3,4$) (common
for all lattice elements) is the basic variable of the construction. So we have
\begin{equation}
{\bf e}_{\mu} = \sum_A E^A_{\mu} {\bf f}_A
\end{equation}
(Everywhere space - time indices are denoted by Greek letters contrary to the
tetrad ones.) Contrary to \cite{TELEPARALLEL} we imply here that all of the
vectors ${\bf e}_{\mu}$ for the simplicial case and vectors ${\bf e}_{\mu}, \mu
= 0, 1,2,3,4$ in the hypercubic case are independent. Also we imply that all
vertices of lattice elements are ordered in some way. The other vectors in
hypercubic case are defined in such a way, that opposite sides of the lattice
element are parallel to each other. The hypercubic lattice is periodic and the
position of the starting point of each lattice element is always denoted by
${\bf e}_0$.

 Variables $E^A_{\mu}$ represent  translations from the fixed point (which will be called later by
 the center of the lattice element) to vertices of the lattice element.
Metric (or vierbein) is implied to be constant inside each lattice element. Its
derivative is singular and is concentrated on the sides\footnote{That are 3 -
dimensional subsimplices in simplicial case and cubes in hypercubic case.} of
the lattice elements. The translation along a path, which consists of the
pieces that belong to different lattice elements is defined as the sum of the
correspondent translations inside those lattice elements.

Here shift of the center of lattice element by a vector $v^A$ causes change in
basic variables: $E^A_{\mu} \rightarrow E^A_{\mu} + v^A$, which is treated as
gauge transformation with respect to the translational gauge group. It
represents the translation of the given lattice element within the
correspondent local map.

Our construction is the special case of Weitzenbock space with singular
torsion, which is of $\delta$ - functional type and is concentrated on the
sides\footnote{In \cite{TELEPARALLEL} we considered approximate expression for
the torsion, which tends to the correct one in the naive continuum limit.
Therefore in \cite{TELEPARALLEL} lattice torsion was attached to the bones,
that are 2 - dimensional subsimplices in simplicial case and plaquettes in
hypercubic case.}  of the lattice elements. We do not discuss here how torsion
is expressed through variables $E^A_{\mu}$ as it follows from the correspondent
expressions of the next section in the limit of vanishing $O(4)$ connection.

\section{Discretization of Riemann-Cartan space.}

We construct the discretization of Riemann-Cartan space as the generalization
of the construction considered in the previous section. Namely, again our space
is composed of flat pieces (simplices or hypercubes). Now in addition to the
translational connection, which is defined by the set of variables $E^A_{\mu}$,
each shift from one lattice element to another is accompanied by the rotation
in the four - dimensional tangent space. In other words, there is the $O(4)$
connection, which is singular and is concentrated on the sides of lattice
elements. We denote by $U_{IJ}$ the $O(4)$ matrix, which is attached to the
side that is common for the lattice elements $I$ and $J$.

The constructed Riemann - Cartan space has singular connection. In this case
definitions of curvature and torsion become ambiguous. Therefore we must fix
one of the definitions in order to calculate them.

Connection is singular on the sides of lattice elements. $O(4)$ curvature is
concentrated on the bones\footnote{That are 2 - dimensional subsimplices  in
simplicial case and plaquettes in hypercubic case.}. We choose the following
integral equation as a definition of $O(4)$ curvature.
\begin{eqnarray}
&&{\rm exp} (\int_{y\in{\Sigma}}\Omega(z,y)
R_{\mu\nu}(y)\Omega^+(z,y)dy^{\mu}\wedge dy^{\nu}) = P {\rm exp}
(\int^z_{z\in \partial \Sigma} \omega_{\mu} dx^{\mu})\nonumber\\
&& {\rm at} |\Sigma|\rightarrow 0
\end{eqnarray}

Here $\omega_{\mu}$ is $O(4)$ connection\footnote{$U_{IJ}=P {\rm exp} (\int
\omega_{\mu} dx^{\mu})$, where integral is over the path of minimal length,
which connects centers of lattice elements $I$ and $J$.}. $\Sigma$ is a small
surface, that crosses the given bone, and $ |\Sigma|$ is its area. $\partial
\Sigma$ is the boundary of $\Sigma$. Its orientation corresponds to orientation
of $\Sigma$. $\Omega(z,y) = P {\rm exp} (\int^y_z \omega_{\mu} dx^{\mu})$ is
the parallel transport along the path that connects a fixed point on $\partial
\Sigma$ with the point $y$. We choose this path in such a way, that it is
winding around the given bone in the same direction as $\partial \Sigma$. The
integral in the right hand side is over the path $\partial \Sigma$, which
begins and ends at the point $z$.

It is worth mentioning, that the given definition does not contradict with the
conventional one in case of smooth connection. And it gives us the possibility
to calculate curvature in the case of the constructed singular piecewize -
linear manyfold.

 Let us fix the
given lattice element. Inside it lattice curvature is equal to
\begin{equation}
R_{\mu\nu B}^A(y) = \frac{1}{D!}\sum_{b\in bones}\int_{x\in{b}}\epsilon_{\mu
\nu \rho \sigma}dx^{\rho}\wedge dx^{\sigma}\delta^{(4)}(y-x) [{\rm Log}
\Pi_{i}U^b_{I_iI_{i+1}}]^A_B,\label{RB1}
\end{equation}
Here the sum is over the bones that belong to the given lattice element. The
integral is over the surface of the bone. The product of the rotation matrices,
which are encountered, while going around the given bone $b$ is denoted as
$\Pi_{i}U^b_{I_iI_{i+1}}$. Here we imply, that this closed path begins within
the given lattice element and has the minimal lattice length.

Now let us calculate torsion, which is concentrated on the sides of lattice
elements. The torsion field $T_{\mu \nu}^A$ is defined by the integral equation
\begin{equation}
 \int_{y\in{\Sigma}}\Omega^A_B(z,y)
T^B_{\mu\nu}(y)dy^{\mu}\wedge dy^{\nu} = \int_{\partial \Sigma}\Omega^A_B(z,y)
b^B_{\mu}(y)dy^{\mu}
\end{equation}
Here $b^A_{\mu}(x)$ is the field of inverse vierbein, which is expressed
through our variables $E^A_{\mu}$ inside each lattice element if the given
parametrization of the lattice element is chosen.

This equation is satisfied with the following expression (which is valid within
the lattice element $\bf I$):
\begin{eqnarray}
T^A_{\mu\nu}(y) & = & \sum_{s\in sides} [\int^{\bf J^s}_{x\in{s}}\frac{[U_{\bf
 IJ^s}]^A_B b^B_{[\mu}(x)\epsilon_{\nu] \tau \rho \sigma}}{D!}dx^{\tau}\wedge dx^{\rho}\wedge
dx^{\sigma}\delta^{(4)}(y-x)\nonumber\\
&& - \int^{\bf I}_{x\in{s}}\frac{b^A_{[\mu}(x)\epsilon_{\nu ]\tau \rho
\sigma}}{D!} dx^{\tau}\wedge dx^{\rho}\wedge
dx^{\sigma}\delta^{(4)}(y-x)]\label{TT}
\end{eqnarray}
Here the first integral in the sum is over the given side $s$ seing from the
neighbor lattice element $\bf J^s$ (the side $s$ is common for $\bf I$ and $\bf
J^s$). (We imply that  in (\ref{TT}) the given lattice element and all its
neighbors have the common parametrization.)

\subsection{Simplicial case}

In order to simplify the last expression let us concentrate our attention on
the given side $s$ (that is $3$ - dimensional simplex). Let us denote its
vertices as $u_i$ ($i = 0, 1,2, 3$). We fix the basis, which is composed of
vectors $h_i= (u_0 u_i), i = 1,2,3$ and the vector $h_0$ of unity length
orthogonal to $h_i, i = 1,2,3$. Then we denote ${\cal E}^{sA}_i = E^A_{u_i} -
E^A_{u_0}$. The vector of unity length orthogonal to ${\cal E}^s_i$ (that is
the image of $h_0$) is denoted as ${\cal E}^{sA}_0 = \frac{\epsilon^{ABCD}{\cal
E}^{sB}_1 {\cal E}^{sC}_2 {\cal E}^{sD}_3}{|\epsilon^{ABCD}{\cal E}^{sB}_1
{\cal E}^{sC}_2 {\cal E}^{sD}_3|}$ . Then nonzero elements of the term (in the
sum of (\ref{TT})) correspondent to the given side $s$  enter the expressions
for $T^A_{i 0}(y)$:
\begin{equation}
 \frac{1}{D}[\int_{x\in
s} ([U_{\bf
 IJ^s}]^A_B {\cal E}^{sB}_i ({\bf J^s})
-   {\cal E}^{sA}_i ({\bf I})) d^3 x \delta^{(4)}(y-x)],\label{TTT}
\end{equation}
where ${\cal E} ({\bf I})$ is calculated inside  lattice element $\bf I$ while
${\cal E} ({\bf J^s})$ is calculated inside lattice element $\bf J^s$.

Next, in tetrad components we have
\begin{eqnarray}
&&T^A_{C D}(y)  =  -  T^A_{DC}(y)   \nonumber\\
&&=\frac{1}{D}\sum_{s\in sides}{\cal E}^{si}_{[C}({\bf I}){\cal
E}^{s0}_{D]}({\bf I}) \int_{x\in s} ([U_{\bf
 IJ^s}]^A_B {\cal E}^B_i ({\bf J^s})  -   {\cal E}^{sA}_i ({\bf I})) d^3 x \delta^{(4)}(y-x)],\label{TTT1}
\end{eqnarray}
where ${\cal E}_A^{si} ({\bf I})$ are the elements of the inverse matrix
$[{\cal E}^s({\bf I})] ^{-1}$.

In order to calculate in tetrad components the expression for the curvature we
fix the following basis within the lattice element $\bf I$, which corresponds
to the bone $b$. Let $\{v^b_0, v^b_1, v^b_2, v^b_3, v^b_4\}$ be the vertices of
the given lattice element, and $\{v^b_0, v^b_1, v^b_2\}$ be the vertices of the
given bone $b$. Then our basis consists of vectors $h^i = (v^b_0 v^b_i), i =
1,2,3,4$. Also we denote ${\cal F}^{bA}_i({\bf I}) = E^A_{v^b_i} -
E^A_{v^b_0}$. Then
\begin{equation}
R_{C F B}^A(y) = \sum_{b\in bones}{\cal F}^{b 3}_{[C}({\bf I}) {\cal F}^{b
4}_{F]}({\bf I})\frac{2}{D!}\int_{x\in{b}}d^2x \delta^{(4)}(y-x)
\Omega^{bA}_B({\bf I}) ,\label{RB}
\end{equation}
where ${\cal F}^{b i}_A({\bf I})$ are the elements of the inverse matrix
$[{\cal F}^b({\bf I})]^{-1}$ and the product of $O(4)$ martices around the
given bone is denoted as $\Omega^{bA}_B({\bf I}) = [{\rm Log}
\Pi_{i}U^b_{I_iI_{i+1}}]^A_B$.

\subsection{Hypercubic case}

Let us define inside each lattice element the following variables: ${\cal
E}^A_{\mu} = E^A_{\mu} - E^A_0, \mu = 1,2,3,4$. Also we denote by ${\cal
E}^{\mu}_A$ elements of the inverse matrix ${\cal E}^{-1}$ In tetrad components
we have:
\begin{equation}
R_{C F B}^A(y) = {\cal E}^{\mu}_{C} {\cal E}^{\nu}_{F}\frac{1}{D!}\sum_{b\in
bones}\int_{x\in{b}}\epsilon_{\mu \nu \rho \sigma}dx^{\rho}\wedge
dx^{\sigma}\delta^{(4)}(y-x) [{\rm Log} \Pi_{i}U^b_{I_iI_{i+1}}]^A_B,\label{RB}
\end{equation}

Torsion is expressed as
\begin{eqnarray}
T^A_{C F}(y) & = & \frac{{\cal E}^{\mu}_{{\bf I}C} {\cal E}^{\nu}_{{\bf
I}F}}{D!} \sum_{s\in sides} \int_{x\in{s}} dx^{\tau}\wedge dx^{\rho}\wedge
dx^{\sigma}\delta^{(4)}(y-x) \nonumber\\&&([U_{\bf
 IJ^s}]^A_B {\cal E}^B_{{\bf J^s}[\mu}\epsilon_{\nu] \tau \rho \sigma} - {\cal E}^A_{{\bf I}
[\mu}\epsilon_{\nu ]\tau \rho \sigma})
\end{eqnarray}
Here new subscript index has appeared in $\cal E$ in order to make difference
between this variable calculated in the given lattice element $\bf I$ and in
its neighbor $\bf J^s$.

\subsection{Gauge transformations}
Now gauge transformations are represented by translations and $O(4)$ rotations
of the lattice elements, that result in the following change in basic
variables:
\begin{equation}
E^A_{\mu} \rightarrow \Theta^A_B E^B_{\mu} + v^A,\label{trans}
\end{equation}
where $\Theta$ is the rotation matrix and $v$ is the vector that represents
translation.

 We must notice, that torsion and curvature
defined above become undefined on the bones and links respectively, where
several sides (bones) intersect each other. It could be possible, in principle,
to define these variables on such objects. Say, if we define torsion on the
bone as the weighted sum of the torsions on the sides, incident on this bone,
then we come to the definition of \cite{TELEPARALLEL} (after $O(4)$ connection
is set to zero). We do not discuss here such a procedure for the curvature, but
imply, that it could be made if necessary.

\section{The action}
\subsection{Simplicial lattice}
At this stage we consider Riemann-Cartan space only as the way to put
Riemannian space on the lattice. Therefore, our action should be constructed in
order to make the model close to the model based on Riemannian space with the
action (\ref{S1}) for some values of coupling constants. That's why we consider
the action in the form
\begin{eqnarray}
S &=& \int \{ \alpha (R_{AB} R_{AB} - \frac{1}{3}R^2) + \beta R^2 - \gamma
m_p^2 R + \lambda m_P^4\} |E| d^4 x \nonumber\\&& +\,  \delta m_P^2 \int
T^A_{BC} T^A_{BC} |E| d^4 x, \label{S2}
\end{eqnarray}
where the second term is added in order to suppress torsion at $\delta
\rightarrow\infty$.

In order to obtain compact form of the expression for the action on the
simplicial lattice we define  lattice tetrad components of curvature and
torsion inside the given lattice element $\bf I$:
\begin{eqnarray}
&& {\bf R}_{C F B}^A({\bf I}) = \sum_{b\in bones}{\cal F}^{b 3}_{[C}({\bf I})
{\cal F}^{b 4}_{F]}({\bf
I}) \Omega^{bA}_B({\bf I})\nonumber\\
&& {\bf R}_{F B}({\bf I}) = {\bf R}_{A F B}^A({\bf I}) \nonumber\\
&& {\bf R}({\bf I}) = {\bf R}_{A A} ({\bf I})\nonumber\\
&& {\bf T}^A_{C D}({\bf I}) = \sum_{s\in sides}{\cal E}^{si}_{[C}({\bf I}){\cal
E}^{s0}_{D]}({\bf I}) [ [U_{\bf
 IJ^s}]^A_B {\cal E}^{sB}_i ({\bf J^s})  -   {\cal E}^{sA}_i ({\bf I}) ]
\end{eqnarray}

Then the action has the form:
\begin{eqnarray}
S & = & \sum_{\bf I\in simplices}  \{ \bar{\alpha} {\bf R}_{FB}({\bf I}){\bf
R}_{FB}({\bf I}) + (\bar{\beta}-\frac{1}{3}\bar{\alpha}){\bf R}({\bf I})^2 +
\bar{\lambda} m_P^4 \nonumber\\ && - \bar{\gamma} m_p^2 {\bf R}({\bf I}) +\,
\bar{\delta} m_P^2 {\bf T}^A_{BC}({\bf I}){\bf T}^A_{BC}({\bf I})\} |{\bf
E}({\bf I})|, \label{S3}
\end{eqnarray}
where $|{\bf E}({\bf I})|$ is the volume of the lattice element $\bf I$.

Here we introduce lattice couplings $\bar{\alpha}, \bar{\beta}, \bar{\delta},
\bar{\lambda}, \bar{\gamma}$ that differ from the original ones by the factors,
which are formally infinite and come from delta - functions in expressions for
torsion and curvature. We assume here, that a certain regularization is made,
which makes these factors finite. Our supposition is that after the
renormalization each physical quantity may be expressed through physical
couplings, which differ from the bare ones (both lattice and continuum), and
the infinity encountered here is absorbed into the renormalization factors.

It is easy to understand, that (\ref{trans}) is the symmetry of the action. So,
we have  lattice model with direct manifestation of Poincare gauge invariance.

\subsection{Hypercubic lattice}

In order to write down the needed lattice formula, we first drop to the dual
lattice. Then our rotation matrices are attached to  links while the inverse
vierbein is attached to sites. Let us denote by $U_{\mu}(x)$ the matrix
correspondent to the link, which begins at the site $x$ and points to the
direction $\mu$ ($\mu = \pm 4, \pm 3, \pm 2, \pm1$).   We denote by
$\Omega_{\mu \nu}(x) = U_{\mu}(x) ... $ the product of link matrices along the
boundary of the plaquette, which is placed in the $(\mu \nu)$ plane. The
inverse vierbein, which is attached to the site $x$, is denoted as ${\cal
E}^A_{\mu} = E^A_{\mu} - E^A_0, \mu = 1,2,3,4$. For negative values of $\mu$ we
define  ${\cal E}^A_{-\mu} = - {\cal E}^A_{\mu}$. The inverse matrix for
positive values of indices is denoted by ${\cal E}^{\mu}_A(x) = \{{\cal
E}(x)^{-1}\}^{\mu}_A$. We also expand this definition to negative values of
indices: ${\cal E}^{\mu}_A(x) = {\rm sign}(\mu)\{[{\cal
E}(x)]^{-1}\}^{|\mu|}_A$. We shall denote by $\Delta x_{\mu}$ the shift on the
lattice by one step in the $\mu$ -th direction ($\Delta x_{-\mu} = - \Delta
x_{\mu} $). So, $x+\Delta x_{\mu}$ is the site which is obtained via the shift
from the site $x$ by one lattice spacing in the direction $\mu$ while $x-\Delta
x_{\mu}$ is obtained by the shift in the opposite direction. Thus, $U_{-\mu}(x)
= U^{-1}_{\mu}(x-\Delta x_{\mu})$. Next, we define $\Delta_{\mu} {\cal
E}_{\nu}(x) = U_{\mu}(x){\cal E}_{\nu}(x+\Delta x_{\mu}) - {\cal E}_{\nu}(x)$.
Everywhere we imply summation over the repeated indices. The summation over
space - time indices $\mu,\nu,...$ is implied over $\pm 4, \pm 3, \pm 2, \pm1$.

Further we introduce lattice tetrad components of torsion and curvature:
\begin{eqnarray}
&& {\bf R}_{C F B}^A(x) = {\cal E}^{\mu}_C {\cal E}^{\nu}_F [\Omega_{\mu \nu}]^A_B \nonumber\\
&& {\bf R}_{F B}(x) = {\bf R}_{A F B}^A (x)\nonumber\\
&& {\bf R}(x) = {\bf R}_{A A} (x)\nonumber\\
&& {\bf T}^A_{C F}(x) = \Delta_{[\mu} {\cal E}^A_{\nu]} {\cal E}^{\mu}_C {\cal
E}^{\nu}_F
\end{eqnarray}

Then the action has the form
\begin{eqnarray}
S &=& \sum_{x\in sites}  \{ \bar{\alpha} {\bf R}_{FB}(x){\bf R}_{FB}(x) +
(\bar{\beta}-\frac{1}{3}\bar{\alpha}){\bf R}(x)^2 + \bar{\lambda} m_P^4
\nonumber\\&& - \bar{\gamma} m_p^2 {\bf R}(x) +\, \bar{\delta} m_P^2 {\bf
T}^A_{BC}(x){\bf T}^A_{BC}(x)\} |{\bf E}(x)|, \label{S31}
\end{eqnarray}
where  $|{\bf E}(x)| = {\rm det}{\cal E} $ is the volume of the lattice element
correspondent to the site $x$. Again, lattice coupling constants contain
infinities discussed above.

\section{Problems with measure over discretized geometries}

It is already a traditional point of view, that measure over Riemannian
geometries should be defined in such a way, that it corresponds to a certain
metrics on the space of geometries. In \cite{REGGE2D} we briefly remind how
such a measure is constructed in case of finite dimensional space. It is often
implied that this procedure could be extended to the infinite - dimensional
space of Riemannian geometries. If so,  measure over discretized spaces should
be constructed in order to reproduce the correct measure over continuum
geometries in the limit of infinite number of lattice elements. While moving in
this direction the so - called Lund - Regge measure over Regge skeletons was
suggested, which corresponds to the metric, which is induced on the space of
Regge skeletons by the given  metric on space of continuum geometries.
Unfortunately, it appears, that in dimensions greater, than two, it is almost
impossible to use this measure in real computer simulations as it is
essentially nonlocal. The possible solution was suggested in two dimensions in
\cite{REGGE2D}, but it could not be extended to higher dimensions.

Moreover, below we show, that it is not straightforward, that Lund - Regge
measure corresponds to the given metric in the limit of large number of lattice
elements. We consider finite dimensional analogue of the situation, where
definitely there is no such a correspondence. Namely, let $\Omega$ be the set
in the coordinate plane $(xy)$: $\Omega = \{(x,y):0.5 < x < 1, |y| < 1\}$. Our
aim is to calculate integral $\int_{\Omega}  dx dy f(x,y)$. Let us now consider
the sequence of curves
\begin{equation}
y(x) = {\rm sin} (\frac{\pi n}{x})\label{C}
\end{equation}
correspondent to integer numbers $n$. Each curve approximates $\Omega$. It is
placed within it more and more dense when $n$ tends to infinity. Therefore, it
is natural to suppose, that it is possible to represent integral over $\Omega$
as the limit of integrals over our curves at $n \rightarrow \infty$.
\begin{equation}
\int_{\Omega}  dx dy f(x,y) = {\rm lim}_{n \rightarrow \infty} \int_{0.5}^1 dx
f(x, {\rm sin} (\frac{\pi n}{x})) \lambda(x)\label{IL}
\end{equation}

Here $\lambda(x) dx$ is the measure on the curve to be defined. This situation
is the simplification of the problem of interest, when the given space of
continuum geometries is approximated by the space of Regge skeletons. $\Omega$
is the analogue of the space of continuum geometries while each curve (\ref{C})
is the analogue of a piecewize - flat space with varying link lengths.

Let us suppose, that the measure over $\Omega$ is defined in accordance with
the norm, that is  $(\delta x)^2 + (\delta y)^2$. The induced norm on the $ n $
- th curve is $ [(\frac{\pi n {\rm cos} (\frac{\pi n}{x})}{x^2})^2 + 1](\delta
x)^2$. Then logic, which has led to the definition of Lund-Regge measure over
discretized Riemannian geometries (seen as Regge manyfolds)\cite{AMBJORN},
would lead us to the expression $\lambda(x) = \sqrt{[(\frac{\pi n {\rm cos}
(\frac{\pi n}{x})}{x^2})^2 + 1]}$, which is obviously incorrect!

The problem is that we missed that different pieces of the curve (\ref{C}) are
distributed with variable density within $\Omega$. So, the correct answer would
be
\begin{equation} \lambda(x) = \sqrt{[(\frac{\pi n {\rm cos} (\frac{\pi
n}{x})}{x^2})^2 + 1]}\frac{1}{\rho(x)},\label{LAM}
\end{equation}
where $\rho$ is the density of the curve inside $\Omega$, which can be
estimated as
\begin{equation}
\rho(x) \sim l_{\epsilon}(x) \sim \frac{\pi n}{x^2},\label{RH}
\end{equation}
where $l_{\epsilon}(x)$ is the length of the curve inside the $\epsilon$ -
vicinity of $x$ (for $\frac{1}{n} << \epsilon << 1$). This can also be proved
by the following sequence of expressions:
\begin{eqnarray}
\int_{\Omega}  dx dy f(x,y) & = & \pi \, {\rm lim}_{n \rightarrow \infty}
\int_{0.5}^1 dx f(x, {\rm sin} (\frac{\pi n}{x})) \sqrt{[({\rm cos}^2
(\frac{\pi n}{x}) + \frac{x^4}{\pi^2 n^2}}\nonumber\\
& = & \pi \, {\rm lim}_{n \rightarrow \infty} \int_{0.5}^1 dx f(x, {\rm sin}
(\frac{\pi n}{x}))  |{\rm cos} (\frac{\pi n}{x})| \nonumber\\
& = & {\rm lim}_{n \rightarrow \infty} \sum_{k = 0, ..., n-1} \int_{-1}^{1} dy
f(\frac{1}{2-\frac{k}{n}}, y) \frac{1}{(2-\frac{k}{n})^2 n}
\end{eqnarray}
Here the first expression is the equation (\ref{IL}) with $\lambda$ substituted
in the form (\ref{LAM}), where $\rho$ given by (\ref{RH}). Constant of
proportionality is chosen to be equal to $\pi$. The last expression is simply
the discretization of integral over $x$ with points $x_k =
\frac{1}{2-\frac{k}{n}}$ ($k = 0, ..., n-1$).

So, the lesson of this example is that if we try to calculate  measure over
discretized geometries using the induced norm, then we may come to the
incorrect result. In particluar, Lund-Regge measure used in Regge
discretizations of Riemannian spaces could be incorrect and further
investigation should be made in order to determine there the analogue of
density $\rho$ of (\ref{LAM}).

\section{Another way to construct measure over discretized geometries}

From the previous section we know, that it may not be possible to calculate
lattice measure using the expression for the continuum norm. Therefore, we
should find another solution. And this solution is right in front of us.
Namely, let us remember QCD on the lattice, which is known to work perfectly.
In this model there are two kinds of fields.

1. There are quarks and leptons. The correspondent measure over Grassmann
variables is well defined and unique.

2. There is the gauge field. The correspondent measure on the lattice is unique
as it is completely defined via symmetry properties: this is the local measure
invariant under gauge transformations.

So, our solution comes easily. We must find symmetry properties of the
continuum measure, which  make the measure unique when it is transferred to
lattice.

In our case there are two fields: $O(4)$ connection and the translational
connection. So, it is natural to use measures, which are invariant under
lattice realization of the gauge transformation. Our choice of measure is
measure, which is simultaneously  invariant under lattice gauge transformations
and is local.

Each piecewise linear manyfold  described above is itself a Riemann - Cartan
space. Let the given discretization (with varying $E$ and $U$) be denoted as
$\cal M$. Then, let $G_{\cal M}$ be the set of correspondent independent
variables $\{E^A_{\mu}({\bf I}); U_{\bf IJ}\}$. Gauge transformation
corresponds to the shift of each lattice element by the vector $v^A ({\bf I})$
and its rotation $\Theta_{\bf I} \in O(4)$. This transformation acts as
$\{E^A_{\mu}({\bf I}); U_{\bf IJ}\}\rightarrow \{\Theta_{\bf I}E_{\mu}({\bf
I})+v({\bf I})); \Theta_{\bf I}U_{\bf IJ}\Theta^T_{\bf J}\}$.

Here by locality of lattice measure we understand the following. The whole
measure should be represented as a product over the sides of lattice elements
 and over the links, that connect centers of lattice elements with their vertices, of
 measures over the matrices $U$ and vectors $E$ correspondingly:
\begin{equation}
D_{\cal M} (E;U) = \Pi_{\bf I} \Pi_{\mu}D E^A_{\mu}({\bf I}) \Pi_{\bf I, J}
DU_{\bf IJ},
\end{equation}
We call the lattice measure local if inside each lattice element $DE^A_{\mu}$
for the given $\mu$ depends upon $E^A_{\mu}$ only,  and $DU_{\bf IJ}$ for the
given $\bf I, J$ depends upon $U_{\bf I J}$ only. It is easy to understand,
that this requirement together with gauge invariance fixes the only choice of
$DE^A_{\mu}$ and $DU_{\nu}$: $D E^A_{\mu} = \Pi_{A,\mu} d E^A_{\mu}$, while
$DU$ is the invariant measure on $O(4)$.

We must mention, that another locality principle may be formulated. Say, we may
thought that the measure is local if $DE^A_{\mu}$ may depend upon $E^A_{\nu}$
with $\nu \ne \mu$ but may not depend upon $E^A_{\mu}$ from another lattice
element. Then gauge invariance does not fix measure precisely. However, our
choice is more strong requirement, mentioned above, which gives us opportunity
to fix the only local and gauge invariant measure. Future investigation must
show is this choice correct or not.

\section{Discussion and conclusions.}

In this paper we suggest to choose the simplest possible way to quantize pure
gravity using Euclidean path integral formalism. The resulting model is well -
known in literature. It deals with Riemannian space, and its only dynamical
variable is the field of vierbein (or, metrics). After making this "minimal"
choice we encounter the following:

1. The model is renormalizable. This means, that it is sensible, at least on
the level of perturbation expansion.

2. There exists the region of coupling constants, where the model is asymptotic
free. Fortunately, it is this region, where the Euclidean action is bounded
below. So, the model can be naturally explored using Euclidean path integral
formalism.

3. If we choose bare couplings in such a way, that at low enough energy scale
renormalized $\lambda$ and $\gamma$ vanish, then at this scale classical
Einstein equations give global minimum to Euclidean action. So, classical limit
is achieved in the absence of matter.

4. If we add matter to the given model in a traditional way, then classical
Newtonian limit cannot be achieved. Nevertheless, in principle, there exists
the way to overcome this difficulty. Namely, according to original ideas of
Einstein and Infeld, it is possible to consider point - like matter objects as
the singularities of space - time. Then classical solutions around such
singularities would give rise to classical Newtonian potential.

5. The model suffer from loss of unitarity. However, at $\lambda = \gamma = 0$
perturbative ghost disappears. And the only problem is the high derivative
action. There exists the possibility to represent the theory in such a way,
that only first derivatives enter the action. Namely, Riemannian space can be
considered as the limit of Riemann - Cartan space with vanishing torsion. We
consider Poincare connection as the basic variable and do not require vanishing
of torsion from the very beginning. Instead we add to the action the term, that
forces torsion to be close to zero at high enough value of the new coupling
constant. The resulting action contains only first derivatives of Poincare
connection and reproduces the original one, when torsion vanishes dynamically.

6. In order to give sense to Euclidean path integral formalism one should put
the model onto the lattice and choose the way to determine measure over
dynamical variables. It is well - known, that if we try to construct measure
over Riemannian geometries in such a way, that it corresponds to a certain
simple choice of metric on the space of geometries, then we would get nonlocal
and rather complicated measure over dynamical variables in Regge discretization
of the model. This measure is in fact so complicated, that real numerical
simulations seem to be almost impossible. Moreover, it may contain such a
factor, that we have no idea how to calculate it at all.

7. We suggest to overcome difficulty with  measure in discretized gravity as
follows. We again use  Riemann - Cartan space instead of Riemannian one. Our
dynamical variable is Poincare connection.

8. Next, we construct manifestly gauge invariant discretization of Riemann -
Cartan space, which may use rectangular or simplicial lattices. Poincare Gauge
transformation naturally acts on the dynamical variables of the constructed
lattice model. We use gauge invariance together with the lattice locality
principle in order to determine measure over the dynamical variables. Such a
measure is shown to exist. Moreover, our definition of locality fixes only one
gauge invariant measure.

9. Finally, we have manifestly gauge invariant discretization, local measure,
 and the action, which does not contain second derivatives. The model,
therefore, is expected to be useful for numerical simulations.

The author is grateful to S.Kofman for numerous discussions of various
theoretical problems and to A.I.Veselov for kind support. This work was partly
supported by RFBR grants 03-02-16941, 05-02-16306, and 04-02-16079, by Federal
Program of the Russian Ministry of Industry, Science and Technology No
40.052.1.1.1112.

\end{document}